\documentstyle[epsfig,epsf,psfig,preprint,aps]{revtex}

\begin{document} 

\date{\today} 
\title{ Existence and transitions properties of three--deuteron 
muonic molecule ($3d\,2e^-\,\mu^-$) } 
\author{V. B. Belyaev and  D. E. Monakhov } 
\address{Joint Institute for Nuclear Research, Dubna, 141980, Russia} 
\author{S. A. Sofianos} 
\address{Physics Department, University of South Africa, P.O. Box 392, 
Pretoria 0003, South Africa} 
\author{W. Sandhas} 
\address{Physikalisches Institut, Universit\"{a}t Bonn, 53115 Bonn, 
Germany} 
 
\maketitle 
 
\begin{abstract} 
We calculated the energy and the size of the  three-deuteron muonic molecule 
$(3d \, 2e^- \mu^-) = {\rm D}_3 \mu$. It turns out that this system possesses 
two equilibrium positions, one at distances typical for muonic molecules 
and a second one at the usual molecular size. We show, moreover, that the 
fusion probability of the three deuterons is considerably enhanced due to 
the existence of a  $^6$Li$^{*}$ threshold resonance. Our estimates indicate 
that this probability is considerably higher than the decay rate of the 
competing Auger transition.\\\\ 
\end{abstract} 
\newpage
\section{Introduction} 
\label{Intro} 
The physics of two-atomic muonic molecules,  as  $pp \mu$, 
$dd \mu$ or $dt \mu$, has a long and intense history since  
the prediction of such molecules in the forties and their discovery in the 
fifties \cite{breun}. Unfortunately, attempts to produce energy for practical 
use by muon-catalyzed fusion in the $dt \mu$ case, the most promising 
candidate, were up till now unsuccessful. However, the corresponding  
investigations revealed some peculiar aspects of the molecular and nuclear  
physics of such systems. Muonic molecules, e.g., offer the possibility
of measuring the strong interaction between charged nuclei at very low
energies $(E \sim {\rm eV})$ which, without the Coulomb shielding by $\mu^-$,
is not accessible to experiment under normal laboratory conditions.
 
An important aspect of these systems, pointed out in 
\cite{bel1,bel2,bel3,bel4}, concerns the formation of threshold 
resonances of the nuclei involved, which  leads to a considerable 
enhancement of their fusion probability. Such 
resonances occur in a number of systems, as $dt \mu$, $t\,^3{\rm He}
 \mu$ or $d\,^6{\rm Li} \mu$. In the $dt \mu$ case~\cite{breun}, 
for example, the near-threshold state $^5 {\rm He}(3/2^+)$  
increases the cross section for the nuclear transition 
\begin{equation} 
\label{1x} 
	dt \mu \longrightarrow ^5 {\rm He}(3/2^+)
        \mu \longrightarrow ^4{\rm He} +\mu + n 
\end{equation} 
\noindent 
by four orders of magnitude as compared to reactions where no threshold  
resonances occur.\\

In the present work we investigate an analogous three-atomic situation,
namely, the system ($3d\,2e^-\,\mu^-={\rm D}_3\mu$) in which 
one of the electrons in the deuterium molecule $(3d\,3e^-) = {\rm D}_3$ 
is replaced by a muon. Fig. 1. shows that three deuterons can form two $^6 Li^*$
resonances at energies very close to the $3d$ threshold energy,
$E^{3d}_{th}=25.32$\,MeV, i.e., to the energy just needed to disintegrate 
$^6$Li into three deuterons.  This indicates that in ${\rm D}_{3 \mu}$  
three-body fusion can take place enhanced by the formation of   
the $ (J^\pi,T) = (3^-,0)$ resonance, 
\begin{equation} 
	 3d\mu \longrightarrow {}^6{\rm Li^*}(3^-,0)+\mu\,. 
\end{equation} 
It is interesting to note that the ${\rm D}_3$ molecule and its 
${\rm D}^+_3$ ion  are stable and 
have  quite peculiar spectra \cite{ten}. The  ${\rm D}^+_3$ 
ion can be produced in the collision 
\begin{equation} 
\label{2x} 
	{\rm D}_2 + {\rm D}^+_2 \longrightarrow {\rm D}^+_3 +{\rm D} 
\end{equation} 
\noindent 
with a high formation probability of the order  of 
$\lambda \sim 10^9\,{\rm cm}^{-3}\,{\rm sec}^{-1}$. Thus, it can be 
easily produced under laboratory conditions \cite{oka}. We also mention that  
the hydrogen analog ${\rm H}^+_3$ of  the deuterium ion ${\rm D}^+_3$, 
discovered already at the beginning of this century by J.J. Thompson \cite{tom}, 
plays an important role in the atmosphere of Jupiter and, in general, in  
astrochemistry where it protonates the otherwise 
chemically inactive CO molecule \cite{miller}, 
\begin{equation} 
\label{3x} 
	{\rm H}^+_3 + {\rm CO} \longrightarrow {\rm H}_2 + {\rm HCO}^+. 
\end{equation} 
It is, therefore, expected that a sufficient number of ${\rm D}_{3 \mu}$ 
molecules can be produced via the ${\rm D}^+_3$ production and a  
subsequent $\mu^-$  capture, so that there is 
enough material for the 3d fusion process. 
 
In Sect.~II we calculate the energy and the size of the ${\rm D}_3 \mu$ 
molecule using a Born-Oppenheimer-type anzaz for the wave function.  
An estimation of the nuclear transition rate and, in particular, of the  
enhancement factor due to the intermediate $^6{\rm Li}^{*}$ resonance 
formation  is obtained in Sect. III. In order to compare this rate 
with the magnitude of the conventional atomic decay of this molecule,  
we estimate in Sect.~IV the corresponding Auger transition probability. 
Our conclusions are drawn in Sect. V. Finally  some technical details 
concerning the Auger transition are presented in the Appendix.  

\section{Energy and size of ${\rm D}_3 \mu$} 
We assume the motion of  the heavy particles, the deuterons, to be
of adiabatic character and therefore their kinetic energy in the total 
Hamiltonian 
\begin{eqnarray} 
\label{Hmad} 
	\hat H = - \frac{1}{2 m_e} (  \Delta_{\mbox{\boldmath$\rho$}_1} 
	+  \Delta_{\mbox{\boldmath$\rho$}_2} ) 
	 - \frac{1}{2 m_{\mu} } \Delta_{\bf r}  
	+ \frac{1}{| \mbox{\boldmath$\rho$}_1 - \mbox{\boldmath$\rho$}_2|} 
	+ \frac{1}{ | \bf{r} - \mbox{\boldmath$\rho$}_1|} +  
	\frac{1}{ | \bf{r} - \mbox{\boldmath$\rho$}_2 | } 
	\nonumber \\  \nonumber \\ 
	- \sum_{i=1}^{3} \frac{1}{ | \bf{r_i} - \mbox{\boldmath$\rho$}_1 | } 
	-\sum_{i=1}^{3} \frac{1}{ | \bf{r_i} - \mbox{\boldmath$\rho$}_2 | } 
	- \sum_{i=1}^{3}  
	\frac{1}{ | \bf r_i -  r | } + \sum_{i \ne j }^{} \frac{1}{ |  
	\bf{r_i} - \bf{r_j} | }  
\end{eqnarray} 
can be omitted. Here  $\mbox{\boldmath$\varrho$}_1$ , 
$\mbox{\boldmath$\varrho$}_2$,  and {\bf r}
are the coordinates of the electrons and the muon in the  center-of-mass
(CM) system (see Fig. 2) of the three deuterons. The ${\bf r}_i$ denote the 
coordinates of the deuterons.\\
 
For the ground state of the $D_3 \mu$ molecule we use the anzatz 
\begin{equation} 
\label{5x} 
	\Psi = \psi (R,\chi,\theta)\, 
	\left( {\rm e}^{-\displaystyle \alpha /2 ({\bf r} - \nu {\bf r}_1 )^2} +
	{\rm e}^{-\displaystyle \alpha /2 ({\bf r} - \nu {\bf r}_2 )^2} +
	{\rm e}^{-\displaystyle \alpha /2  ({\bf r} - \nu {\bf r}_3 )^2} 
	\right)
	{\rm e}^{\displaystyle  
	 - \kappa /2 \rho_1^2 - \kappa / 2 \rho_2^2} 
\end{equation} 
where in the three-deuteron wave function $\psi(R,\chi,\theta)$ 
the following hyperspherical variables were chosen instead of 
the position vectors {\bf r}$_i$,
\begin{equation} 
\label{xy} 
        R^2 = x^2 +y^2,\qquad \tan{\chi}=\frac{y}{x},\qquad  
	\cos{\theta}=\frac{{\bf x\cdot y}}{x y},  
	\qquad \Omega\equiv(\chi,\theta) 
\end{equation} 
where
\begin{equation} 
	{\bf x } =\sqrt{\frac{1}{2}}  ({\bf r}_1 - {\bf r}_2), \qquad 
	{\bf y}=\sqrt{\frac{2}{3}} \left({\bf r}_3 - 
	\frac{( {\bf r}_1+{\bf r}_2 )}{2} \right). 
\end{equation} \\
Employing the anzatz (\ref{5x}) in a minimization procedure in which the 
parameters $\alpha$, $\nu$  and $\kappa$ at fixed values of
{\boldmath $r_i$} are varied, we obtain the corresponding energy 
surface $U (R, \chi, \theta)$, the effective interaction of the deuterons.
Due to its variational origin, this potential will provide an upper bound to the 
exact eigenvalue of $\hat H$ at any pair of angles $\chi$ and $\theta$. 
As can be seen in Figs. 3 and 4,  the minimum of this function at
different values of $R$ is reached for $\chi_0 = \pi/4$ and 
$\theta_0 = \pi/2$, i.e., for an  equilateral position of the three deuterons. 
Therefore, the ground state for the three-deuterons should appear essentially 
in this symmetric configuration and thus the potential 
$U_0(R)=U(R,\chi_0,\theta_0)$  can be  employed to  calculate 
$\psi(R,\chi,\theta)$. This is supported by the fact that, due to the 
almost classical motion of the heavy particles, the ground state energy will 
be close to the minimum of the  potential surface shown in Fig. 3 and 4.
The dependence on  $\alpha$, $\nu$ and $\kappa$ on $R$ is shown in Fig. 5 
while in Fig. 6 we plot  $U_0(R)$ versus the hyperradius $R$. As can be seen,
 $U_0(R)$ has two minima, the first one at  ``muonic''  distances and the 
second one at atomic distances. The later minimum, obviously, 
corresponds to the case where one of the deuterons in 
 a ${\rm D}_2$ molecule, is replaced by the ${\rm dd}\mu$ bound system.

The eigenvalue equation for the 3d states reads
\begin{equation} 
\label{Shr} 
	\left( -\frac{1}{2 m_d}\frac{1}{R^5}\frac{\partial} 
	{\partial R }R^5\frac{\partial}{\partial R }+ 
	\frac{1}{2m_d} \frac{\Lambda^2}{R^2} + U_0(R)  
	\right )\psi(R,\Omega) =  
        {\cal E}_3 \psi(R,\Omega) \; , 
\end{equation} 
with
\begin{equation} 
        \Lambda^2 =\frac{4}{\sin^2{\chi}}
			\left (
	\frac{\partial}{\partial\chi}\sin^2{\chi} 
	\frac{\partial}{\partial\chi}+\frac{1}{\sin{\theta}}
	\frac{\partial}{\partial\theta}\sin{\theta}\frac{\partial}
	{\partial\theta}
			\right )
\end{equation}
Due to the $\Omega$-independence of $U_0(R)$,  the solutions of (\ref{Shr})
 can be written as
\begin{equation}
	\psi (r, \Omega) = X(R) {\cal Y}_{\cal L} (\Omega),
\end{equation}
where the hyperspherical harmonics ${\cal Y}_{\cal L} (\Omega) $ satisfy 
the eigenvalue equation
\begin{equation}
	\Lambda^2 {\cal Y}_{\cal L} = {\cal L}({\cal L} +1) {\cal Y}_{\cal L}
\end{equation}
of the grand angular momentum operator $\Lambda$. The quantum number
$ {\cal L} = n+l_1+l_2+3/2$ is composed of the angular momenta $l_1$ and $l_2$
corresponding to the Jacobi coordinates ${\bf x}$ and ${\bf y}$,  while $n$
denotes the degree of the relevant Jacobi polynomial.\\    

In a  wide range of $R$ $(0.1 \le R \le R_0)$ and for the quantum number 
${ \cal L}=5/2$ (relevant in our context) we have
\begin{equation}
	\frac{ {\cal L ( L } +1)}{2 m_d R^2} \ll |U_0(R)|.
\end{equation}
Equation (\ref{Shr}) thus reduces to
\begin{equation} 
\label{Shr1} 
	\left( -\frac{1}{2 m_d}\frac{1}{R^5}\frac{\partial}{\partial R } 
	R^5\frac{\partial}{\partial R }+U_0(R) \right )X(R) =  
	{\cal E}_3 X(R)\,.   
\end{equation} 
Having in mind the extremely large width of the barrier between 
the two minima of the potential  $U_0(R)$ one can consider 
the states bound by the short--range minimum as physically stable with 
respect to a decay into the second--minimum states. In this area the 
potential can be approximated fairly well by 
\begin{equation} 
	U_0(R)=a+b(R-{\bar R})^2  \,.
\label{uoappr}
\end{equation} 
\noindent 
The  constants $a$ and $b$ were found to be (we use the units $ e=\hbar=m_\mu =1 $)
\begin{equation}  
	a = 0.31, \qquad b = 0.01, \qquad  {\bar R} = 4.45 \, .
\end{equation}   
Equation (14) can then be easily solved with the 
results for the energy and the size of ${\rm D}_3 \mu$ being 
\begin{equation} 
\label{6x} 
	{\cal E}_3= - 1.74\,{\rm keV}\,, \qquad  R_0 = 11.0 \, 10^{-   
	11}\, {\rm cm}\,. 
\end{equation} 
Both values are of the same order of magnitude as the ones of the
corresponding two-atomic muonic molecule (${\cal E}_2\sim -2.7 
\div -2.8 $\,keV,  $R_0\sim 5\,10^{-11}$\,cm)~\cite{breun}.\\
 
Apart from effects due to the $^6$Li resonance formation  (see forth section 
III) comparable nuclear transition rates  should  be expected in both systems.
%
\section{Nuclear Transitions} 
According to Fig. 1 there are two $^6{\rm Li}^{*}$ resonances with 
energies close to the $3d$ threshold. The lowest one for the
 $(J^\pi$, $I) = (4^-,1)$ quantum numbers is given by
		 $E + i \Gamma = (25 + i4)$\,MeV 
and the  higher one for  $(J^\pi$, $T) = (3^-,\, 0)$ \cite{eisen} which has 
a broad width  $E = 26.6$\,MeV.
Since the isospin of the deuteron is zero, the 3d state can
undergo a fusion reaction only via the latter resonance, and hence
the ${\rm D}_3 \mu$ molecule has to be in a $J^\pi = 3^-$  state. \\ 
  
Following the procedure suggested in~\cite{bel5}, the probability of the   
nuclear fusion is obtained  from the overlap  integral ${\cal M}$
between the molecular and resonance wave  functions, i.e.,  from
\begin{equation} 
	{\rm w}={\cal E}_3 |{\cal M}|^2 \,.
\end{equation}\\ 
In contrast to the calculations of the energy and the size of the ${\rm D}_3 \mu$ 
molecule, where simple approximations
such as (\ref{uoappr}) were sufficient, the calculations of the overlap
integral is  much more delicate. This is due to the fact that the
wave function of the three outgoing charged particles in the 
near-threshold region we are interested in, is highly 
oscillatory. Thus, in order to avoid direct numerical integrations
we choose wave functions that describe the crucial input  for the
${\rm D}_3 \mu$ molecule and the $^6$Li resonance, and allow us  to perform
the integrations analytically.

For this purpose we approximate the potential $U_0(R)$ at short
distances by an analytic function. This is easily achieved by using 
the  effective Coulomb potential $ Z_{eff}/R$. The effective charge
 $Z_{eff}$ can be found by fitting the eigenpotential $U_0(R)$ 
at small distances the result being  $Z_{eff} = 2.8$. In this region the 
wave function is, therefore, given by the regular Coulomb solution 
$F_{\cal L} (R)$.

At large distances  the wave function  decrease exponentially and 
therefore we may  use  a phenomenological  anzaz of the form 
\begin{equation} 
\label{11x} 
	\psi_{mol} = \enspace \frac{N_m}  
	{R^\frac{5}{2}} F_{\cal L} (\eta,kR) 
	\displaystyle\, {\rm e}^{- k R} Y^{J,M}
	(\hat x,\hat y)u_{\cal L}(R) \,, 
\end{equation} 
which describes both the size of the molecule and the Coulomb repulsion of 
the three deuterons at small distances.  $N_m$ is the normalization 
constant,  $\eta = Z_{eff} m_d / k $ is the Sommerfeld parameter, 
and $k = \sqrt{2m_d |{\cal E}_3|}$.
 $ Y^{J,M}(\hat x,\hat y)$ are the bispherical harmonics corresponding to  
the total angular momentum $J$. Since we assume the molecule to be in the 
ground state with negative parity, we have $n = 0$ and for the sum of the 
relative angular momenta $\ell_1 + \ell_2 = 1$. In other words we have
${\cal L} = 5/2$.
 
Due to the repulsion of the deuterons at small distances, the wave function 
(\ref{11x}) is exponentially small in this region. This means, the main 
contribution to the overlap integral comes from distances $R$ 
much larger than the nuclear size $r_0$ and thus we may use the   
asymptotic form of the $^6Li^*$ resonance state which is taken to 
be the asymptotic wave function of three charged particles in the continuum  
\begin{equation} 
\label{12x} 
	\psi_{res} (R, \Omega) = \frac{N_{res}}{R^{5/2}} f^J(R,\omega)  
	Y^{J,M}(\hat x,\hat y), 
\end{equation} 
where 
$$ 
	f^J(R,\omega) = \int \exp[\displaystyle i q R +  
	\displaystyle\frac{i v(\Omega) }{q} \ln( q R)] 
	Y^{J,M}(\hat x, \hat y) \; d \hat x  d \hat y 
$$  
The factor $N_{res}$ represents the normalization in the nuclear volume
 and the angles are denoted by $\Omega\equiv (\omega,\hat x,\hat y)$;
$v(\Omega)$ represents the angular part of the total Coulomb potential 
of the three--body systems written in terms of hyperspherical variables 
while $q =  \sqrt{2m_d|E - E_{th}|}$ with $E$ being the energy of the 
outgoing  particle and $E_{th}$ the threshold energy.\\ 
 
After some tedious calculations, with $\eta(\Omega) \equiv v(\Omega) m_d 
/q\approx \eta \gg 1$,  and $u_{\cal L}(R)=1$, we found that the 
overlap integral, 
\begin{equation} 
\label{13x} 
	{\cal M} = \int  \enspace \psi_{res} (R, \Omega) \psi_{mol} 
	(R,\Omega) \; R ^5dR\, d\Omega\,, 
\end{equation} 
has the behavior
\begin{equation} 
\label{15x} 
	{\cal M} \sim e^{\displaystyle \pi \eta( \Omega_1) 
    (\displaystyle\frac{ i S(t_{-})}{\pi} - 
	\displaystyle\frac{1}{2}  )},   
\end{equation}  
with  
$$ 
	\gamma = \frac{q}{k}\,, \qquad  
	\sigma=\frac{v(\Omega_1)}{Z_{eff}}=1.06\,,   
$$  
$$ 
	S(t)= [ -\ln(t) +\ln(1-t) ] \frac{\gamma}{\sigma}  
	+ \ln(1 -i (\gamma -1) -2 i t)\,, 
$$ 
and  
$$ 
	t_{-}=\frac{\sigma -\gamma}{2 \sigma} - \frac{1}{2 \sigma} 
	\sqrt{\gamma^2 +\sigma^2 - 2 \gamma \sigma [\gamma +i]}\,. 
$$ 
\noindent 
$\Omega_1$ is obtained from the condition
$$ 
	\nabla v(\Omega)=0 
$$  
and turns out to be the same as $\Omega_0 \equiv (\chi_0,\theta_0)$ 
introduced above.  It should be noted that any other reasonable choice 
of $u_{\cal L}(R)$ is not expected  to lead to a significant modification 
of ${\cal M}$.

From~(\ref{15x}) it follows that an enhancement of the transition rate 
is achieved if  
\begin{equation}
\label{cond} 
    {\rm Im\, S}(t_-) < 0\,, \qquad
  	\frac {{\rm |Im\, S}(t_-)|}{\pi} > \frac{1}{2}\,.
\end{equation}  
In Table \ref{tab1} we present the ${\rm Im\,S}(t_-)/\pi$ for certain values  of 
the parameters $\gamma$. Evidently the condition~(\ref{cond}) is satisfied. 
In other words, the usual decreasing Coulomb barrier is replaced  by a 
factor which grows with increasing $\eta$.  
Since in the present case (near threshold reactions) $\eta \gg 1$, a 
considerably enhanced transition of the three deuterons into the 
$^6{\rm Li}^{*}(3^-,0)$ resonance is to be expected. The corresponding
transition probabilities are given in the last column of Table I.  
The above nuclear fusion process is accompanied by a radiative $E1$ 
transition into the $^6{\rm Li}$ ground state yielding an energy of 
about 26\,MeV .
%
\section{Auger transition rate} 
Since the  binding energy of ${\rm D}_3 \mu$ is essentially 
higher than the one obtained for the $dd \mu$-molecule \cite{alex},  
we have also as a possible decay process the Auger transition 
which for the ${\rm D}_3 \mu$ molecule  is given by 
\begin{equation} 
\label{7x} 
	{\rm D} _{3 \mu} \longrightarrow X + e^-\, . 
\end{equation} 
Here $X$ includes  the $dd\mu$ molecule, the deuteron $d$, and
a second (bound) electron $e^-$. To specify the wave function of 
the system $X$ the following  arguments can be used.
In the Auger transition (\ref{7x}) most of the  energy  released
is taken by the outgoing electron. This means that the relative velocity of
the $dd\mu$  molecule  and a nucleus $d$ in the final state, is much smaller as
compare to the velocities of electrons. Furthermore having  in mind 
the small distances between all three deuterons in the initial state, one can 
consider the bound  electron as moving in the Coulomb field of a  ``united''
 atom, formated by the $(dd\mu)^+$ ion  and the nucleus $d$.     
 
The transition rate for this mode of decay is, of course, quite essential 
and is of relevance to the competing nuclear transition. 
The corresponding transition probability is determined from the amplitude
\begin{equation} 
\label{8x} 
	{\cal M}_A = <\psi_f|V_C|\psi_i> , 
\end{equation} 
where $V_C$  is the Coulomb interaction of the outgoing 
electron with the second electron and with the center of mass 
the complex $[dd\mu+d]$. 
The initial state $\psi_i$ is the ${\rm D}_3 \mu$ wave 
function described Sec. II, Eq. (\ref{5x}). The final state $\psi_f$ 
is chosen to be a product of the wave functions of the fragments, namely, 
\begin{equation} 
	\label{9x} 
	\psi_f = \,\psi_{rel} \, \psi_{at} \, \phi_{out} \psi_{dd \mu} 
\end{equation} 
where $\psi_{rel}$  is the  Coulomb wave function for the relative motion 
of the $dd\mu$ molecule and the deuteron $d$, $\psi_{at}$ is the atomic wave 
function of the bound electron in the field of the  $dd\mu+d$ system, 
$\phi_{out}$ is the plane wave function of outgoing electron, and 
$\psi_{dd \mu}$  is the ground state wave function of the $dd \mu$ molecule
which is assumed to be of the form
\begin{equation}
	\psi_{dd \mu} =\sqrt{\frac{\zeta^3}{\pi a^3_\mu}}
                   \, {\rm e}^{-\displaystyle \zeta r_d/a_\mu}
            \psi^{LCAO}_{dd\mu}
\label{psiddm}
\end{equation}

where $\psi^{LCAO}_{dd\mu}$ designates the Linear Combination of Atomic 
Atomic Orbitals (LCAO) wave function,
\begin{equation}
      \psi^{LCAO}_{dd\mu}=\frac{1}{\sqrt{2(1+F(r_d))}} 
		\frac{1}{\sqrt{\pi a^3_\mu}}
	  \left(
      	\,{\rm e}^{-\displaystyle 
	|{\bf r}_d/2-{\bf r}_\mu|/a_\mu}\,
    + \,{\rm e}^{-\displaystyle 
	|{\bf r}_d/2+{\bf r}_\mu|/a_\mu}
		\right)\,,
\end{equation}
with
$$
     F(r_d)=	\left( 
	      1+\frac{r_d}{a_\mu}+\frac{1}{3}
	\left(\frac{r_d}{a_\mu}	\right)^2
	\right)	\, {\rm e}^{-\displaystyle r_d/a_\mu}\,.
$$
$a_\mu$ is the Bohr radius  for the $d\mu$ atom, $r_d$ the distance
between the deuterons, $r_\mu$ is the muon Jacobi coordinate in the  
$dd\mu$ system, and $\zeta=0.023$ is a constant that characterizes 
the size of the $dd\mu$ molecule. Due to the comparatively large  
size  of the  ${\rm D}_3 \mu$ molecule, it is enough to use a $dd\mu$ wave 
function of the form (\ref{psiddm}) which provides the correct 
binding energy and has the correct behavior at large distances.

The width for the Auger transition (\ref{7x}) is defined by
(see the Appendix)
\begin{equation}
	\frac{\Gamma}{2}=\frac{\mu_D e^4}{\hbar^2 16\pi^3}\displaystyle
	\int_0^1 {\rm d}x \frac{|{\cal M}_A(x,\tilde q(x))|^2}{\tilde q(x)}
\label{Gamma}
\end{equation}
where 
$$
\begin{array}{llc}
	&x=\displaystyle\frac{ k}{k_0}\,,\qquad& 
	\tilde q(x)=\sqrt{\displaystyle\frac{\mu_D(1-x^2)}{\mu_e}}\,,\\
      &  k_0= \sqrt{\displaystyle\frac{-2\mu{\cal E}_3}{\hbar^2}}\,,\qquad&
         \mu_D=\displaystyle\frac{ m_d(2\mu_d+m_\mu)}{3m_d+m_\mu}\,.\\
\end{array}
$$
Numerical integration of $|{\cal M}_A|^2$ over all final state momenta permitted 
by the energy-momentum conservation yields the decay rate 
\begin{equation} 
\label{10x} 
	\lambda=\frac{\Gamma}{\hbar} \approx 10^{15} {\rm sec}^{-1} . 
\end{equation} 
%
\section{Conclusions} 
Our calculations show that the effective potential energy for the three--deuterons 
in the ${\rm D}_3 \mu$ 
system  possesses two equilibrium positions the first being at 
a distance characteristic for muonic molecules while the second one is of the same 
order of magnitude as that  of  the usual electronic molecules.
The position and value of the  ``atomic'' minimum of the effective potential 
energy imply the following cluster structure of the ${\rm D}_3\mu$ 
system at this distance
\begin{equation}
	\Psi_{\displaystyle D_3 \mu} \approx \Psi_{\displaystyle D} +
 	\alpha \Psi_{\displaystyle (dd\mu) e } \; ,
\end{equation}
where $\Psi_{D}$  is  the wave function of the hydrogen atom and 
$\Psi_{(dd\mu) e }$ is a hydrogen--like wave function with the  $dd\mu ^+$ ion
in the center of such ``atom''.
One should bear in mind that such a clustering ensures the lowest possible
energy of the system ${\rm D}_3\mu$. Indeed, as can be seen in Fig. 4, this 
is just the case. It is obvious, that such a state of the ${\rm D}_3\mu$ 
system is stable as far as 
electromagnetic transitions are concerned, but it is not stable 
with respect to nuclear fusion in the $dd \mu$ subclusters.

Let's now discuss in brief the  possibility of observing the ${\rm D}_3 \mu$ 
system. One way is to search for the Auger electrons with kinetic energy 
around 1\,keV. A second possibility is, of course, more exotic and 
consists in observing the
products of the triple fusion and the formation of the highly exited state 
(3$^-$,0)  at  26.6\,MeV of the  $^6Li$ nucleus.

 The  enhancement factor of the nuclear transition in the ${\rm D}_3 \mu$ 
molecule, obtained above, should be considered as an indication of the
possible strong influence of the near-threshold nuclear resonance on the 
molecular system.


\centerline{\bf Acknowledgement} 
Financial support from the Foundation for Research Development of South Africa 
is greatly acknowledged. One of us (VBB) expresses his gratitude to  
the Physikalisches Institut der Universit\"at Bonn and  
the University of South Africa for their kind hospitality. 

\newpage  
\appendix  
\section{Evaluation of the Auger Transition}

For the width $\Gamma$ the following definition has been
used:
\begin{equation}
\label{gap1}
    \frac{\Gamma}{2}=\int \pi\frac{{\cal M}_A}{2kq}\delta\left(E-
	\frac{\hbar^2k^2}{2\mu_e}-\frac{\hbar^2q^2}{2\mu_D}\right)
	\frac{{\cal M}_A^*}{2 kq}\frac{{\rm d}^3k}{(2\pi)^3}
	\frac{{\rm d}^3q}{(2\pi)^3}\,.
\end{equation}
The matrix elements for the Auger transition ${\cal M}_A$ (\ref{8x}) 
are given by
\begin{eqnarray}
\nonumber
 {\cal M}_A&=&
	\int{\rm d}^3r_D {\rm d}^3 \rho {\rm d}^3r_d {\rm d}^3r_\mu 
	{\rm d}^3r_e 
	\Psi_{dd\mu}(r_d,r_\mu)\frac{F_0(\eta,qr_D)}{r_D}
	\frac{\sin(k\rho)}{\rho}\frac{2^{3/2}}{\sqrt{\pi a^3_e}}
              \,{\rm e}^{\displaystyle- 2 r_e/a_e}\\
     	&&N_{var}\left( - \frac{2}{\rho} +\frac{1}{| \mbox{\boldmath$\rho$} -{\bf r}_e |} \right)
    	 \, {\rm e}^{\,\displaystyle-\frac{k}{2}\rho^2}\,
	{\rm e}^{\displaystyle -\frac{k}{2}\, r_e^2}
        \, \left( \sum_{i=1}^{3}
 	{\rm e}^{\,-\displaystyle \frac{\alpha}{2}\,({\bf r}_\mu-
	\frac{1}{3}{\bf r}_D- \nu {\bf r}_i)^2}\right) \; X(R),
\end{eqnarray}
where $N_{var}$ is the normalization constant of the variational part 
of the wave function (\ref{5x}) and $r_D$ is the Jacobi coordinate  of the 
third deuteron not belonging to the $dd\mu$ molecule.

After some analytical derivations, one remains with the following
simplified form
\begin{eqnarray}
\nonumber
 	{\cal M}_A&(&k,q)= \frac{16 \; 2^{3/2} \sqrt{\zeta}}{\pi^{7/4}} 
	\left(\frac {a_\mu}{a_e} \right)^{1/2}
	\int\frac{\displaystyle {\rm d}^3r_d {\rm d}^3r_D}
	{\sqrt{\displaystyle 3 \left(1+2 \exp{(-\displaystyle \frac{1}{4} 
	\alpha \nu^2 R^2  )}\right)}}  \frac{ \kappa^{3/2} }
	{ \alpha^{3/4} }\,
	\frac{{\rm e}^{\,-\displaystyle \zeta\,r_d}}{\sqrt{2(1+F(r_d))}}\\
\label{M_fin}
	&&\left(\sum_{i=1}^{3} I_+^{(i)} + I_-^{(i)} \right) 
        \frac{F_0(\eta,q r_D)}{r_D} \,
	\left( - {\cal J}( \kappa / 8 )\left(\frac{a_e}{a_\mu} \right)
	\frac{k}{4 \kappa} {\cal K} (k^2 / \kappa) + 
	{\cal I} \left(\kappa /2,k\,\frac{ a_e}{ a_\mu} \right)
   	\right) \, X(R)
\end{eqnarray}
where 
$$
	R^2=\frac{1}{2}r_d^2+\frac{2}{3}r_D^2\,,
\qquad
I_\pm^{(i)}=\frac{4\pi}{\sqrt{ \alpha}d_{i\pm}}{\rm e}^{\displaystyle
	-\alpha \frac{d_{i\pm}^2}{2}  } {\cal R}\left(\frac{1}
	{\sqrt{\alpha }},\sqrt{\alpha d^2_{i\pm}}\right)\,,  
$$

$$
	{\cal R}(\xi,\sigma)=-\frac{\sqrt{2\pi}}{4}
        \left[(\xi-\sigma)\,{\rm e}^{\displaystyle-(\xi-
	\sigma)^2/2}\,{\rm erfc}
	\left(\frac{\xi-\sigma}{\sqrt 2}\right)-
         (\xi+\sigma)\,{\rm e}^{\displaystyle-(\xi+\sigma)^2/2}
	\,{\rm erfc}\left(\frac{\xi+\sigma}{\sqrt 2}\right)\right]\,,
$$

$$
{\vec d}_{i\pm}=\displaystyle\frac{1}{3}\vec r_D\pm\frac{1}{2} \vec r_d +
\nu \vec r_i \,,
\qquad
	{\cal K} (x)=\sum_{n=1}^\infty\,\frac{(-x)^{n-1}}{(2n-1)!!}\,.
$$

$$
{\cal J }(\alpha)=- \frac{1}{4 \alpha^2}+\sqrt{\frac{\pi}{\alpha^5}}\,
\frac{1/2+\alpha}{4}  \,\displaystyle e^{\displaystyle\frac{1}{4 \alpha}}\, 
\left(1- {\rm erf } \left( \frac{1}{2 \sqrt{\alpha}}\right) \right)
$$

$$
{\cal I }(\alpha,\delta)=\int_0^{\infty} {\rm d}x\,x\,\sin(\delta x) \,
e^{\displaystyle -\alpha\,x^2 } \left( \frac{1}{x} J_1(x,\alpha) +
J_2(x,\alpha) \right)
$$

\begin{eqnarray}
\nonumber
	J_1(x,\alpha)&=&e^{\displaystyle 1/ \alpha}\, \left\{ 
 	\frac{1}{2 \alpha^{3/2}}
	\left[   \gamma(3/2,\alpha(x+ 1/\alpha )^2) -\gamma(3/2,1/
	 \alpha)  \right]\right . 
\\ \nonumber
	&&\left . - \frac{1}{\alpha^2 }\, \left[ e^{\displaystyle - 1/\alpha }- 
	e^{\displaystyle - \alpha(x+ 1/\alpha )^2 } \right] 
 	- \frac{1}{2 \alpha^2}\sqrt{ \frac{ \pi }{ \alpha }}
	\left[ 1 - {\rm erf} \left( \sqrt{\alpha}(x+1/\alpha ) \right)  \right] 
 	\right\}\,,
\end{eqnarray}
and
$$
	J_2(x,\alpha)=e^{\displaystyle 1/ \alpha}\,\frac{1}{2\alpha} \, \left( 
	{\rm e}^{\displaystyle - \alpha(x+ 1/\alpha )^2 } - 
	\sqrt{\frac{\pi}{\alpha}}\left[ 1 - {\rm erf} 
	\left(\sqrt{\alpha}(x+1/\alpha ) \right)  \right]\right)\,.
$$
We note here that in (\ref{M_fin}) the following dimensionless variables 
	$\alpha a_\mu^2  \to \alpha$,
	$\kappa a_e^2  \to \kappa$,
	$k a_\mu  \to k $,
	$q a_\mu  \to q $,
	$r_d / a_\mu  \to r_d $, and
	$r_D / a_\mu  \to r_D $ are used.


\begin{table} 
\caption{ The function ${\rm Im\, S}(t_-)$ for three different values of 
of the parameter $\gamma$ and   $\sigma = 1.06$.} 
\vspace{.5cm} 
\label{tab1} 
\begin{tabular}{cccc} 
$\gamma $ & $\eta(\Omega_0)$ &     $\rm Im S(t_-)/\pi$ & 
 $\rm w
\;\;s^{-1}$\\ 
\hline 
\hline 
$0.857$   & $21.20$ & $-0.5244$ & $5.42\;\;10^{19}$ \\ 
$1.916$   & $9.48$  & $-1.5130$ & $3.37\;\;10^{44}$ \\ 
$2.710$   & $6.70$  & $-2.2590$ & $3.03\;\;10^{50}$ \\ 
\end{tabular} 
\end{table} 
 
 
\vspace*{3cm} 
\begin{figure} 
\epsfig{file=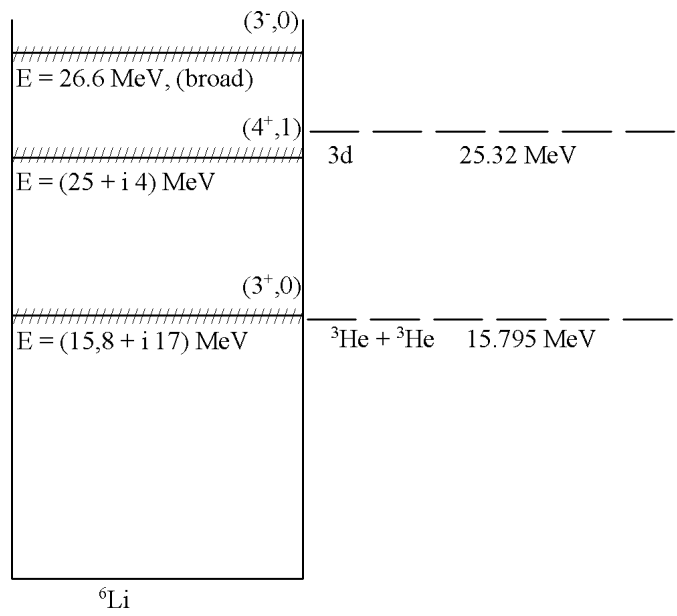,height=10cm}
\caption{The ${}^6{\rm Li}$ nucleus spectrum.} 
\end{figure} 

\newpage 
 
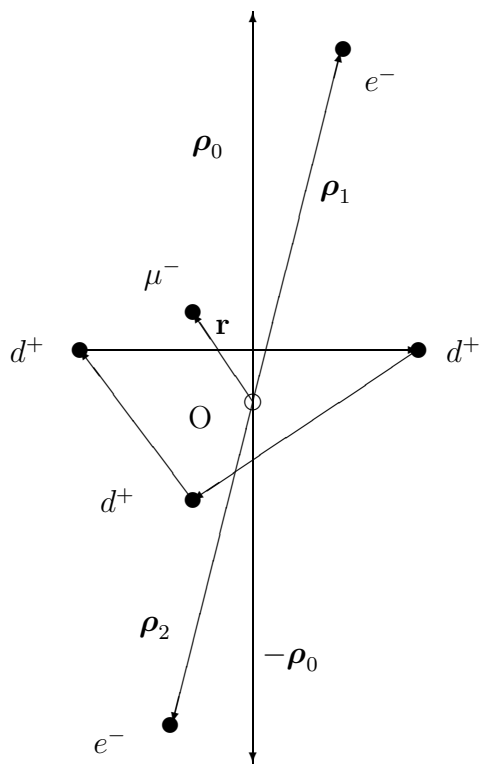
\begin{figure} 
 
\vspace*{3cm} 
\unitlength=1.00mm 
\special{em:linewidth 0.4pt} 
\linethickness{0.4pt} 
\begin{picture}(91.00,120.00) 
\put(40.00,75.00){\circle*{2.00}} 
\put(85.00,75.00){\circle*{2.00}} 
\put(55.00,55.00){\circle*{2.00}} 
\put(55.00,55.00){\vector(-3,4){15.00}} 
\put(40.00,75.00){\vector(1,0){45.00}} 
\put(85.00,75.00){\vector(-3,-2){30.00}} 
\put(63.00,68.00){\circle{2.00}} 
\put(55.00,80.00){\circle*{2.00}} 
\put(63.00,68.00){\vector(-2,3){8.00}} 
\put(63.00,68.00){\vector(1,4){11.67}} 
\put(63.00,68.00){\vector(0,1){52.00}} 
\put(63.00,68.00){\vector(0,-1){48.00}} 
\put(75.00,115.00){\circle*{2.00}} 
\put(63.00,68.00){\vector(-1,-4){10.67}} 
\put(52.00,25.00){\circle*{2.00}} 
\put(57.00,102.00){\makebox(0,0)[cc]{$\mbox{\boldmath$\rho$}_0$}} 
\put(74.00,96.00){\makebox(0,0)[cc]{$\mbox{\boldmath$\rho$}_1$}} 
\put(80.00,111.00){\makebox(0,0)[cc]{$e^-$}} 
\put(91.00,75.00){\makebox(0,0)[cc]{$d^+$}} 
\put(33.00,75.00){\makebox(0,0)[cc]{$d^+$}} 
\put(45.00,55.00){\makebox(0,0)[cc]{$d^+$}} 
\put(51.00,85.00){\makebox(0,0)[cc]{$\mu^-$}} 
\put(59.00,78.00){\makebox(0,0)[cc]{$\bf r$}} 
\put(68.00,34.00){\makebox(0,0)[cc]{$-\mbox{\boldmath$\rho$}_0$}} 
\put(44.00,23.00){\makebox(0,0)[cc]{$e^-$}} 
\put(50.00,38.00){\makebox(0,0)[cc]{$\mbox{\boldmath$\rho$}_2$}} 
\put(56.00,66.00){\makebox(0,0)[cc]{O}} 
\end{picture} 

\caption{Jacobi coordinates for all constituent particles of the 
${\rm D}_3 \mu$ molecule.} 

\end{figure} 
 
%
 
\newpage 
\begin{figure} 
\centerline{\psfig{figure=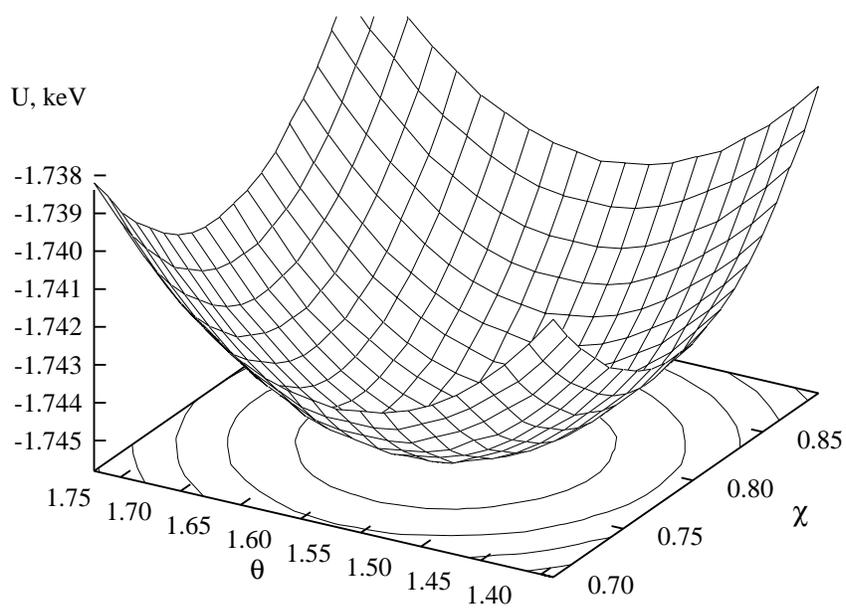,width=15cm,angle=0}} 
\vspace*{7.0cm}
\caption{Two--dimensional cross section of  $U(R,\chi,\theta)$ 
with   ${\rm R}=4.5 \; a_{\mu} $. 
The surface has a minimum at $\theta=\pi/2$ and $\chi=\pi/4$
which corresponds to the equilateral triangle formed by 
 the 3d nuclei.}  
\end{figure} 

 
\newpage 
\begin{figure}  
\centerline{\psfig{figure=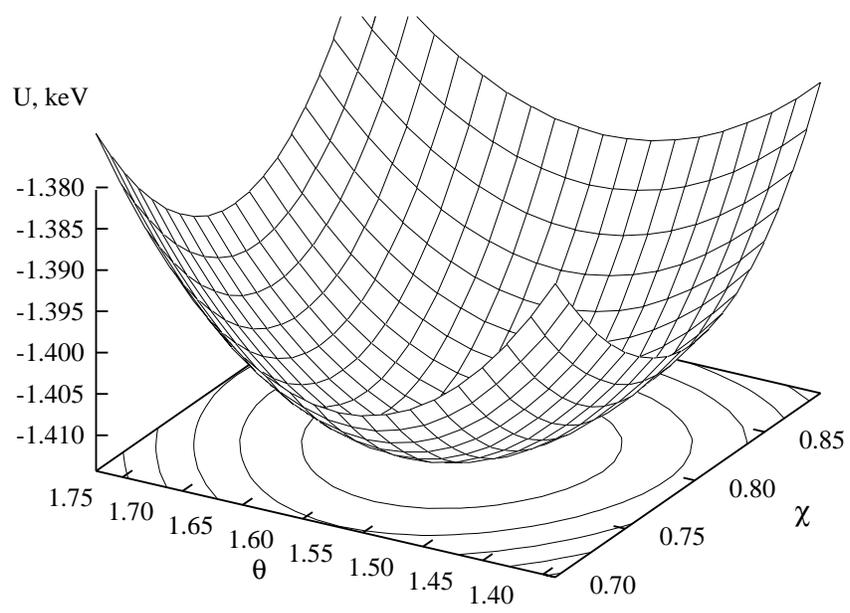,width=15cm,angle=0}} 

\vspace{7.0cm} 
\caption{Same as in Fig. 3 with ${\rm R}=2.5 \;a_{\mu}$.} 

\end{figure} 


\newpage 
\vspace*{8cm} 
\begin{figure}  
\centerline{\psfig{figure=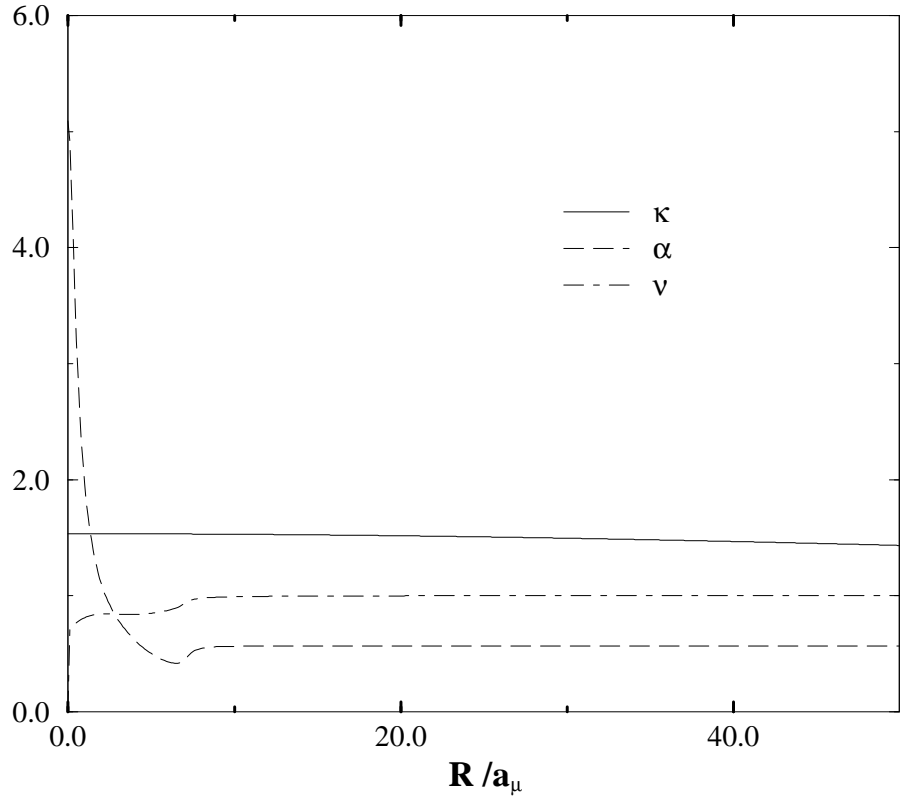,width=15cm,angle=0}} 
\vspace{4cm} 
\caption{The dependence of the parameters $\kappa$, $\alpha$, and $\nu$
 of Eq. (6) on $R$ } 

\end{figure}

 
\newpage 
\vspace*{2.0cm} 
\begin{figure} 
\centerline{\psfig{figure=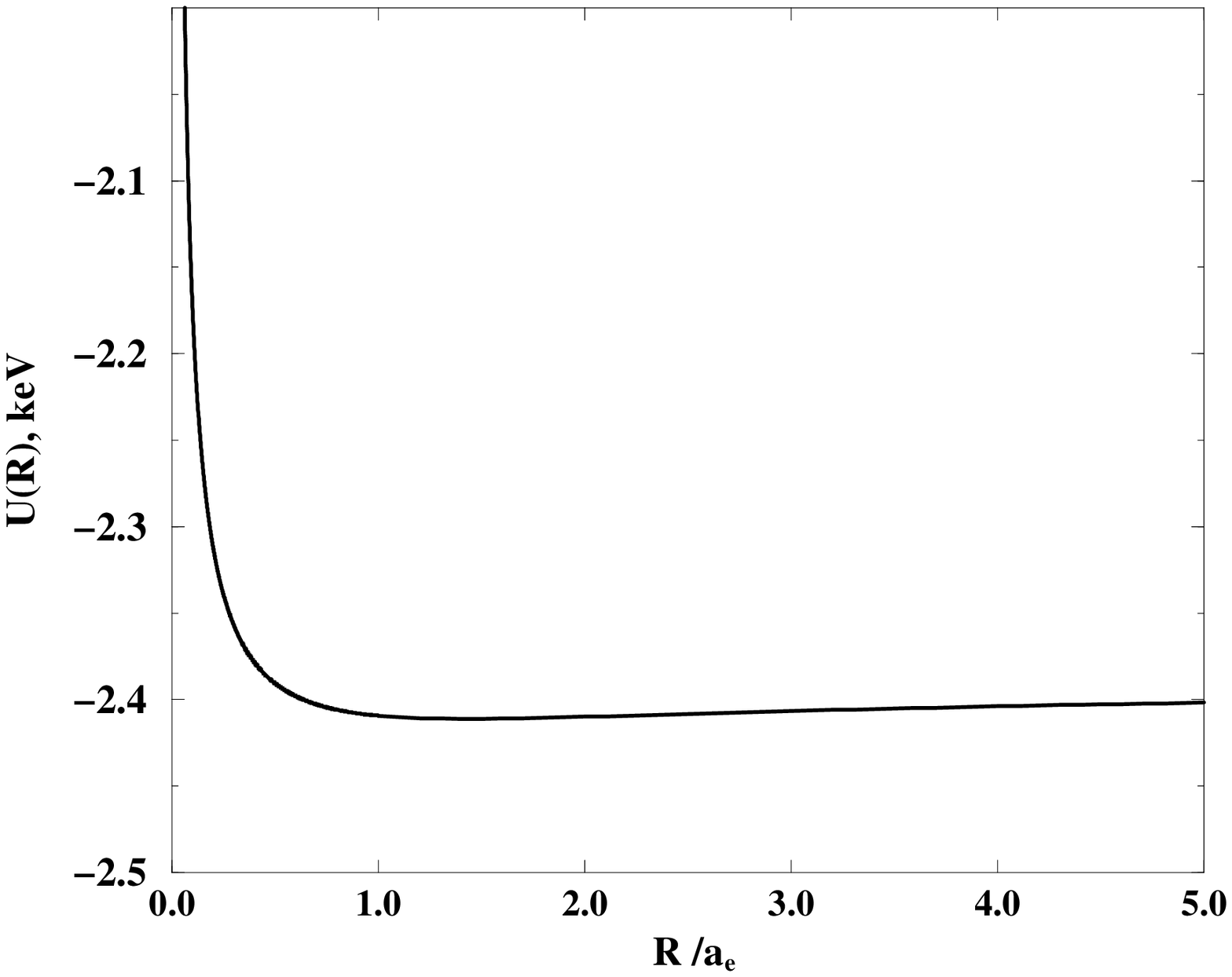,width=11.5cm,angle=0}} 
\vspace{5.5cm} 
\centerline{\psfig{figure=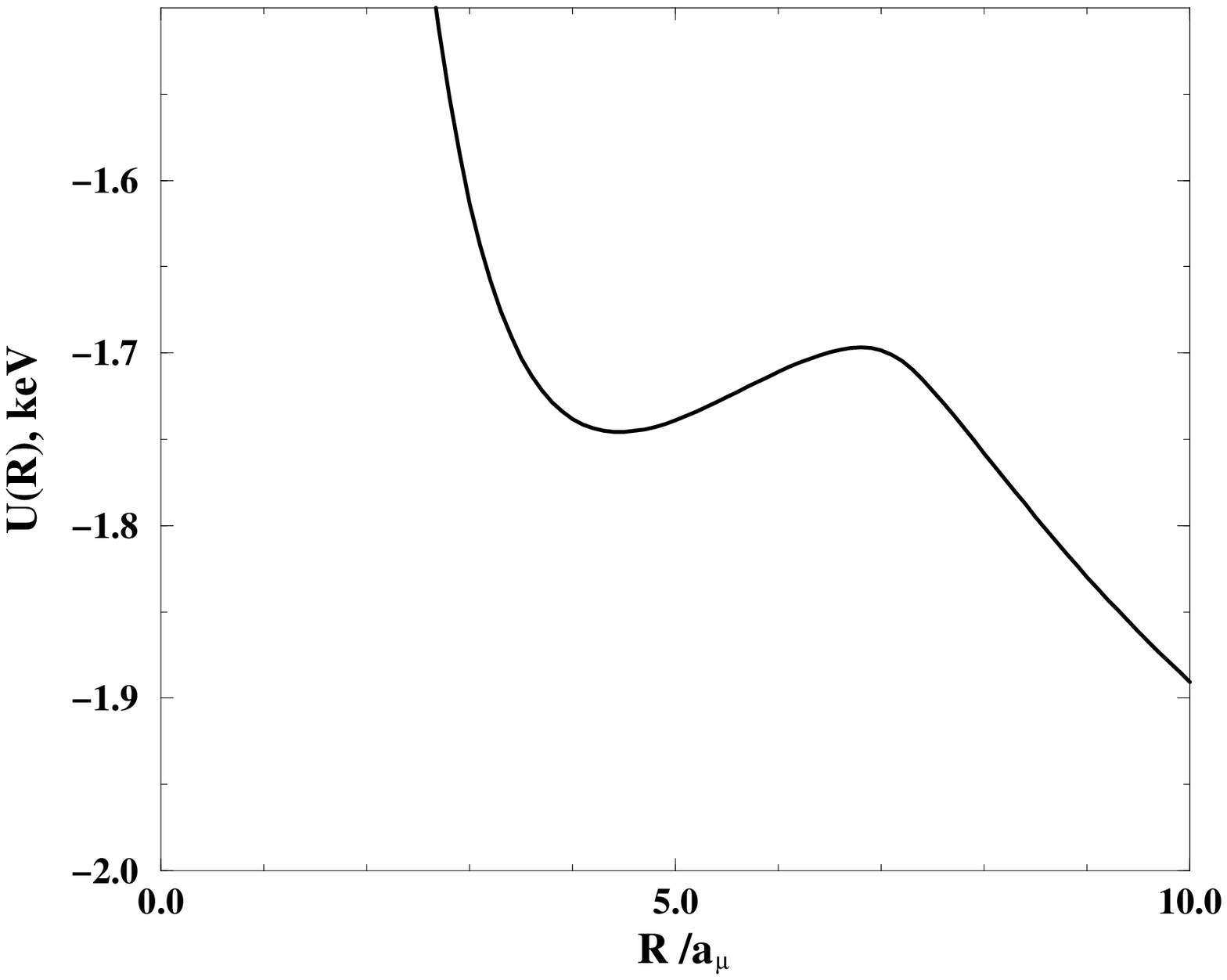,width=11.5cm,angle=0}} 
\vspace{4cm} 
\caption{The $U_0(R)= U(R,\chi_0,\theta_0)$ effective potential.
On the upper figure the potential
is shown at atomic distances while on the lower at mesonic distances. 
The $a_e$  and   $a_\mu$ are the electronic and muonic
Bohr radii respectively.} 
\end{figure} 
 

\begin{references} 
\bibitem{breun} 
	 W. H. Breunlich et al., Ann. Rev., Part. Sci. {\bf 39}, 311 (1989). 
\bibitem{bel1}  
	V. B. Belyaev, M. Decker, H. Fiedeldey, S. A. Rakityansky, 
	W. Sandhas, and S. A. Sofianos, Nucleonica {\bf 40}, 3 (1995). 
\bibitem{bel2}  
	V. B. Belyaev, H. Fiedeldey, and S. A. Sofianos, Phys. At. 
	Nucl. {\bf 56}, 893 (1993). 
\bibitem{bel3}  
	V. B. Belyaev, S. A. Rakityansky, H. Fiedeldey, and S. A. 
	Sofianos, Phys. Rev. A {\bf50}, 305 (1994). 
\bibitem{bel4}  
	V. B. Belyaev, A. K. Motovilov, and W. Sandhas, J. Phys. G 
	{\bf 22}, 1111 ( 1996). 
\bibitem{alex}   
	S. A. Alexander and H. J. Mankhorst, Phys. Rev. A {\bf 38},
	26 (1988).
\bibitem{ten} 
	J. Tennyson, Rep. Progr. in Physics {\bf 58}, 421 (1995). 
\bibitem{oka}  
	T. Oka, Rev. Mod. Phys. {\bf 64}, 1141 (1992). 
\bibitem{tom}  
	J. J. Tompson, Phil. Mag. {\bf 21}, 235 (1911);
   {\bf 21}, 209 (1912); 
\bibitem{miller}  
	S. Miller and J. Tennyson, Chemical Soc. Rev. {\bf 21}, 281 (1992). 
\bibitem{bel5}  
	V. B. Belyaev, A. K. Motovilov, and W. Sandhas, Physics-Doklady
    {\bf 41}, 514 (1996).
\bibitem{eisen}
         F. Eisenberg-Selove Nucl. Phys. {\bf A413}, 1 (1984).  
\end{references}
\end{document}